\documentstyle[12pt,epsf]{article}
\begin{document}
\begin{flushleft}Interner Bericht \\
DESY-Zeuthen 96--07  \\
October 1996
\end{flushleft}

\vspace{2cm}

{\LARGE{Study of Z$'$ Couplings to Leptons and Quarks at NLC}
\footnote{To appear in {\it Proceedings of 1996 DPF/DPB Summer Study
on New Directions for High Energy Physics (Snowmass 96)}, Snowmass,
CO, 25 June -- 12 July 1996.}
}

\vspace{1cm}

\begin{center}
{Sabine Riemann\\ 
{\it DESY--Institut f\"ur Hochenergiephysik,
 Platanenallee 6, D--15738 Zeuthen, Germany}}
\end{center}

\vspace{1cm}


\begin{abstract} 
\noindent
The discovery of a Z$'$ and the
measurement of Z$'$ couplings to fermions are main tasks of
future colliders.
Here, the possibilities
to measure Z$'f \bar f$ couplings at an NLC operating
below a Z$'$  resonance are studied.
In dependence on the mass
of the Z$'$ and the collider parameters
one will be able to discriminate  between Z$'$ models.
\end{abstract}

\section{Introduction}
%

The Standard Model is the powerful theory to
describe processes observed at existing colliders.
But, most physicists are thoroughly convinced that beyond the Standard Model
surprises are ahead.
A Z$'$ is an extension of the Standard Model 
with clear possibilities of detection and interpretation.
If the centre--of--mass energy of an NLC
is large enough to produce Z$'$ bosons the study of their
properties will be easy.
But even indirect measurements of $e^+e^- \rightarrow f \bar f$
below the Z$'$ production threshold
may give information about its nature.
Besides information about the mass the knowledge of
Z$'$ couplings to fermions is
important for a study of the symmetry breaking expected at
an energy of 1 TeV.

As long as the Z$'$ mass is unknown, a precise and
model-independent measurement of
Z$'f\bar f$ couplings, $a'_f,~v'_f$ is difficult.
One has to analyze directly cross sections and
asymmetries \cite{zefit}.
Nevertheless, all observables are
sensitive to normalized Z$'$ couplings $a_f^N, v_f^N$;
e.g. $a^N_f = a'_f\sqrt{s/(m_{Z'}^2-s)}$  (see \cite{al}).
Only if the centre--of--mass energy is near the production threshold
Z$'$ mass and Z$' f \bar f$ couplings
can be determined with  good accuracy and
the identification of
Z$'$ models becomes possible (see \cite{rizzo_nlc, alsr}).

Here, prospective  measurements  of Z$'$
couplings to fermions
at NLC are studied assuming that the  Z$'$ mass is known
from a discovery at LHC.

In preparation of the design for an NLC different scenarios of basic
parameters are suggested (see \cite{snow_rep}):
\begin{eqnarray}
 \mathrm{(a)} &\sqrt{s} = ~500 \mathrm{GeV} :&  ~50 {fb}^{-1} \nonumber\\
 \mathrm{(b)} &\sqrt{s} = 1000 \mathrm{GeV} :&  100 {fb}^{-1} \label{coll}\\
 \mathrm{(c)} &\sqrt{s} = 1500 \mathrm{GeV} :&  100 {fb}^{-1} \nonumber
\end{eqnarray}
The electron beam will be polarized, $P_{e^-}=80$~\%, the positron beam
is assumed to be unpolarized. Here, case (\ref{coll} a)  and
(\ref{coll} c) are taken into account.
Expectations for (\ref{coll} b) or other scenarios
can be extrapolated.

A collider with the parameters (\ref{coll}) allows the 
measurement of
total cross section, $\sigma_T$, forward-backward asymmetry,
$A_{FB}$, left-right asymmetry, $A_{LR}$ and
forward-backward polarization asymmetry, $A_{LR}^{FB}$ 
with small statistical uncertainties.
The experience of LEP and SLD experiments  encourages
to expect good techniques of quark flavour identification
with high efficencies and purities \cite{d_jackson}.
Nevertheless,
background reactions and misidentification of  final state fermions
can lead to relatively large systematic errors
especially if $q \bar q$ final states are analysed.
The systematic errors can dominate 
and weaken the results for the Z$' f \bar f$ couplings (see also
\cite{rizzo_nlc,alsr})
The influence of systematic errors
on Z$'$ coupling measurements is also considered.

In the following, the measurements of Z$' l \bar l$ couplings
and  Z$' q \bar q$ couplings are  studied separately.

Cuts are applied to approach a realistic situation of measurements.
In case of leptonic
final states an angular acceptance cut of 20$^o$ is taken into account.
Further, the  t-channel exchange in Bhabha scattering
is neglected.
If quark flavours are tagged the fiducial volume depends on the design
of the vertex detector. Prospective designs foresee a selection
of at least two tracks within 25$^o < \theta <155^o$.
For completeness,
the influence of different angular ranges on the
accuracy of the Z$' f \bar f$  coupling measurement is also considered.

In order to reach the full sensitivity to Z$'$ effects, 
a cut on the energy
of photons emitted in the initial state, $\Delta = 1 - s'_{min}/s$, is
applied. At $\sqrt{s}=500$~GeV a radiative return to the Z peak is
avoided choosing e.g. $\Delta = 0.9$.
An uncertainty of 0.5\% is taken into account
for the luminosity measurement.
For numerical studies  the program package ZEFIT/ZFITTER \cite{zefitp,zfitter}
is used.

\section{Z$'$ Couplings to Leptons}


Assuming lepton universality  Z$'$ couplings to the initial and final
state are equal. This clean scenario allows the determination of
Z$'$ couplings to leptons with a good accuracy if the difference
between centre-of-mass energy and Z$'$ mass is not too large.
Taking into account a systematic  error of 0.5\% for all leptonic
observables and an efficiency of lepton identification of 95\%
the Z$'$ couplings can be identified
as demonstrated in figure 1.
It is assumed  that a Z$'$ exists either in the $\chi$
model or in the LR model. Different
Z$'$ masses are assumed.

The observables depend only on products or squares of $a'_f$ and
$v'_f$. Thus, a two-fold ambiguity in the signs of couplings remains.

The sensitivity to Z$'$ effects in $e^+e^-$ annihilation
is reduced if the Z$'$ mass is larger.
A scaling law describes this  (see \cite{al}):
\begin{equation}
\frac{\Delta a'_1}{\Delta a'_2};~ \frac{\Delta v'_1}{\Delta v'_2}
\approx \displaystyle{ \frac{m_{Z'_1}^2-s^2}{m_{Z'_2}^2-s^2} }
\label{scal_lept}
\end{equation}
If $m_{Z'} \ge 6\cdot \sqrt{s}$ the Z$'$ contributions influence
the observables only weakly.
The point $(a'_l, v'_l) = (0, 0)$ in figure 1 cannot be excluded
with ~95\% CL, although 
the existence of a Z$'$ ($\chi$ model) is assumed.
A discrimination between models, even the indirect detection
of a Z$'$ is no longer possible.
Only upper limits on Z$'$ parameters can be derived.

With a higher luminosity the loss of sensitivity may be
compensated,
\begin{equation}
\frac{\Delta a'_1}{\Delta a'_2};~ \frac{\Delta v'_1}{\Delta v'_2} 
\approx   \left(
\frac{{\cal L}_2}{{\cal L}_1} \right)^{1/4}  
\label{eq_lumi}
\end{equation}
Relation (\ref{eq_lumi}) shows that an increase of luminosity
improves the accuracy of the Z$'$ coupling measurement only slowly.
If possible, it is better to go to higher centre--of--mass energies.
An NLC operating
at $\sqrt{s}=1.5$~TeV  with a luminosity of 100~fb$^{-1}$
allows
a clear distinction between Z$'$models up to $m_{Z'}=3$~TeV
analyzing leptonic  observables.
This is shown in figure~2. 
The Z$'$ mass is assumed to be known.

Systematic errors of leptonic observables
are of the magnitude assumed above may be neglected.

\begin{figure}[htbp]
  \begin{center}
   \mbox{\epsfxsize=8cm\epsffile{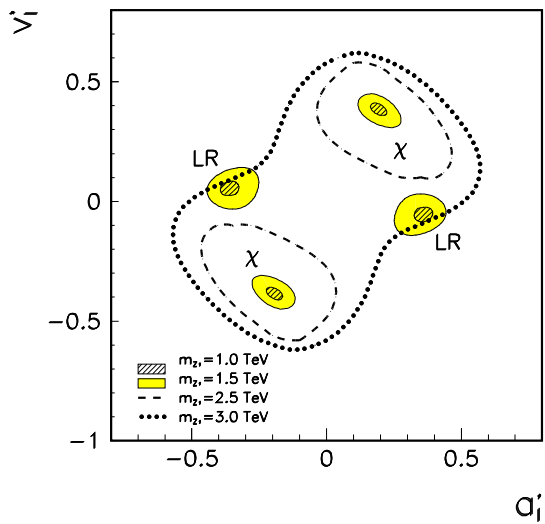}
}
 \end{center}
\caption{95\% CL contours for $a'_l$ and $v'_l$. A Z$'$ is assumed
in the $\chi$ model or in the LR model
with a mass of  $m_{Z'}$=1 TeV (hatched area) and
$m_{Z'}$=1.5 TeV (shaded area).  The dashed (dotted) line limits the
95\% CL bounds  on Z$' l \bar l$ couplings if a Z$'$ with a mass
$m_{Z'}$=2.5 TeV ($m_{Z'}$=3 TeV) exists in the $\chi$ model;
${\cal L}=50$~fb$^{-1}$ and $\sqrt{s} = 500$~GeV.
}
\label{fig:1}
\end{figure}

\begin{figure}[htbp]
  \begin{center}
   \mbox{\epsfxsize=8cm\epsffile{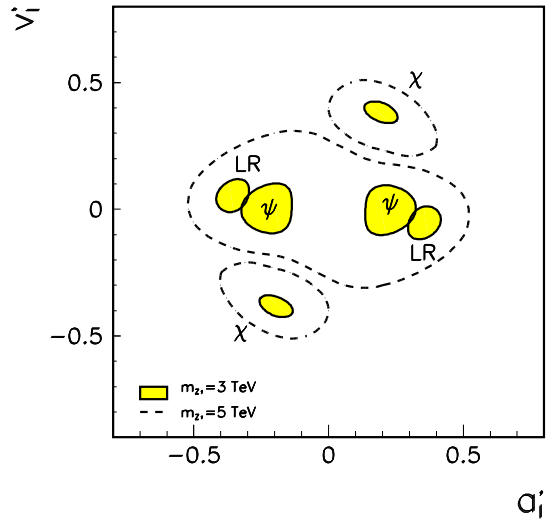}
}
 \end{center}
\caption{Discrimination between $\chi$, $\psi$ and LR model based on
95\% CL contours for $a'_l$ and $v'_l$ for $m_{Z'}$=3~TeV.
For $m_{Z'}= 5$~TeV (dashed line) $\chi$ and LR model can be still
distinguished;
${\cal L}=100$~fb$^{-1}$ and $\sqrt{s} = 1.5$~TeV.
}
\label{fig:2}
\end{figure}

\section{Z$'$ Couplings to Quarks}

At  NLC the derivation of  Z$'$ couplings to quarks
depends on the knowledge of the couplings to electrons.
In particular, if the error range 
of $a'_e$
and $v'_e$ includes $a'_l=v'_l=0$, a simultaneous fit to leptonic
and quarkonic couplings fails. In the following we assume that
an analysis of leptonic observables leads to non-vanishing
Z$  e \bar e$ couplings.

The identification of quark flavors is more complicated than
lepton identification. 
Although very promising designs of a vertex detector for the NLC
let expect efficiencies of 60\% in b--tagging with a purity of
80\%  \cite{d_jackson}, the systematic error of b-quark observables
will not be less than 1\%. This is suggested from extrapolating
the present experience of the SLD collaboration with roughly
150 k hadronic Z$^o$ events  \cite{b_sld}
up to the collider perfomance given in (\ref{coll}).
The systematic
errors limit the accuracy of a $a'_q, v'_q$ determination
substantielly. Improvements due to a higher luminosity can be fully
removed by an imperfect flavor identification.
Figure 3 and 4 demonstrate the influence of  integrated luminosity,
Z$'$ mass and  systematic errors  on the contours for 
Z$' q\bar q$ couplings.
An efficiency of 40\% is assumed
for the determination of Z$' c \bar c$ couplings in
agreement with \cite{d_jackson}.
Both figure 3 and 4 demonstrate that the magnitude of the systematic
errors is important for Z$' q \bar q$ coupling measurements.

\begin{figure}[htbp]
  \begin{center}
   \mbox{\epsfxsize=8cm\epsffile{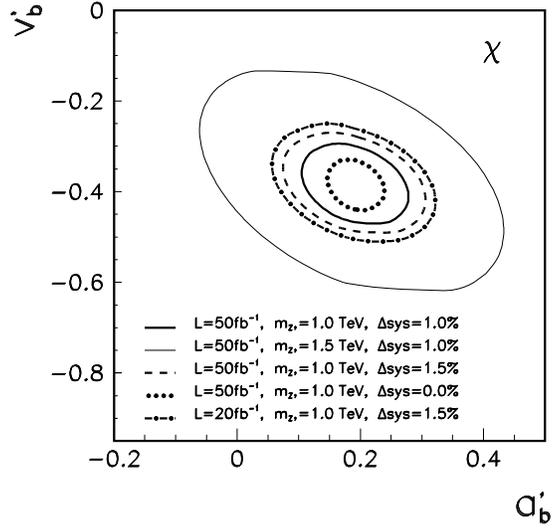}
}
 \end{center}
\caption{Influence of luminosity,  Z$'$ mass, and systematic error
 on contours of Z$' b \bar b$ couplings.
A Z$'$ in the $\chi$ model is assumed.}
\label{fig:3}
\end{figure}

\begin{figure}[htbp]
  \begin{center}
   \mbox{\epsfxsize=8cm\epsffile{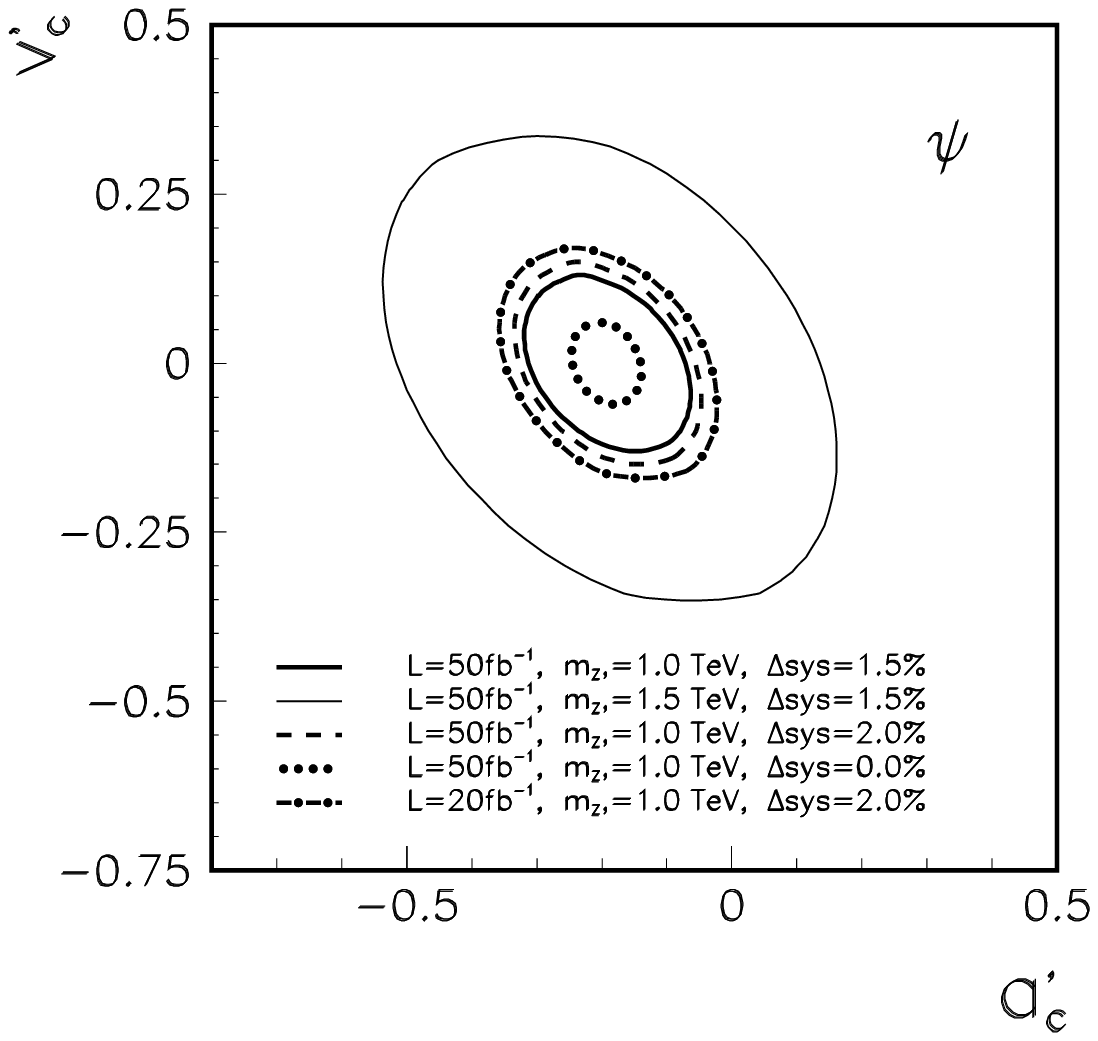}
}
 \end{center}
\caption{Influence of luminosity, Z$'$ mass, systematic error
 on contours of Z$' b \bar b$ couplings.
A Z$'$ in the $\chi$ model is assumed.}
\label{fig:4}
\end{figure}

\begin{figure}[htbp]
  \begin{center}
   \mbox{\epsfxsize=8cm\epsffile{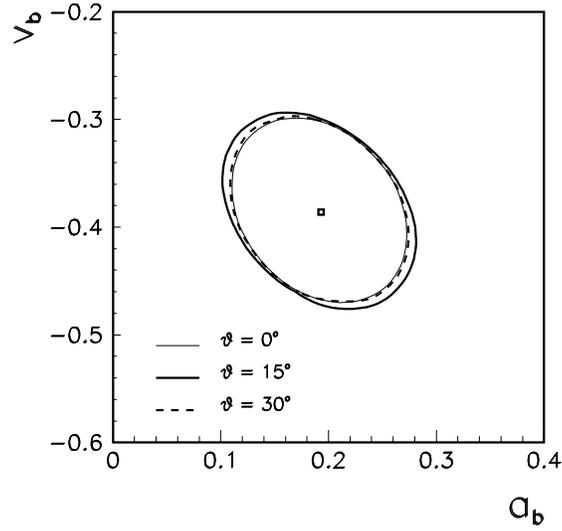}
}
 \end{center}
\caption{95\% CL contours of Z$'$ couplings to b-quarks in dependence
on the fiducial volume of the vertex detector. An ideal 4$\pi$ detector
(thin solid line) is compared with $\cos \theta < 0.96$ (dashed line)
and $\cos \theta < 9.87$ (solid line). 
A systematic error of 1\% for all
b-quark observables is  taken into account.}
\label{fig:5}
\end{figure}

\begin{figure}[htbp]
  \begin{center}
   \mbox{\epsfxsize=8cm\epsffile{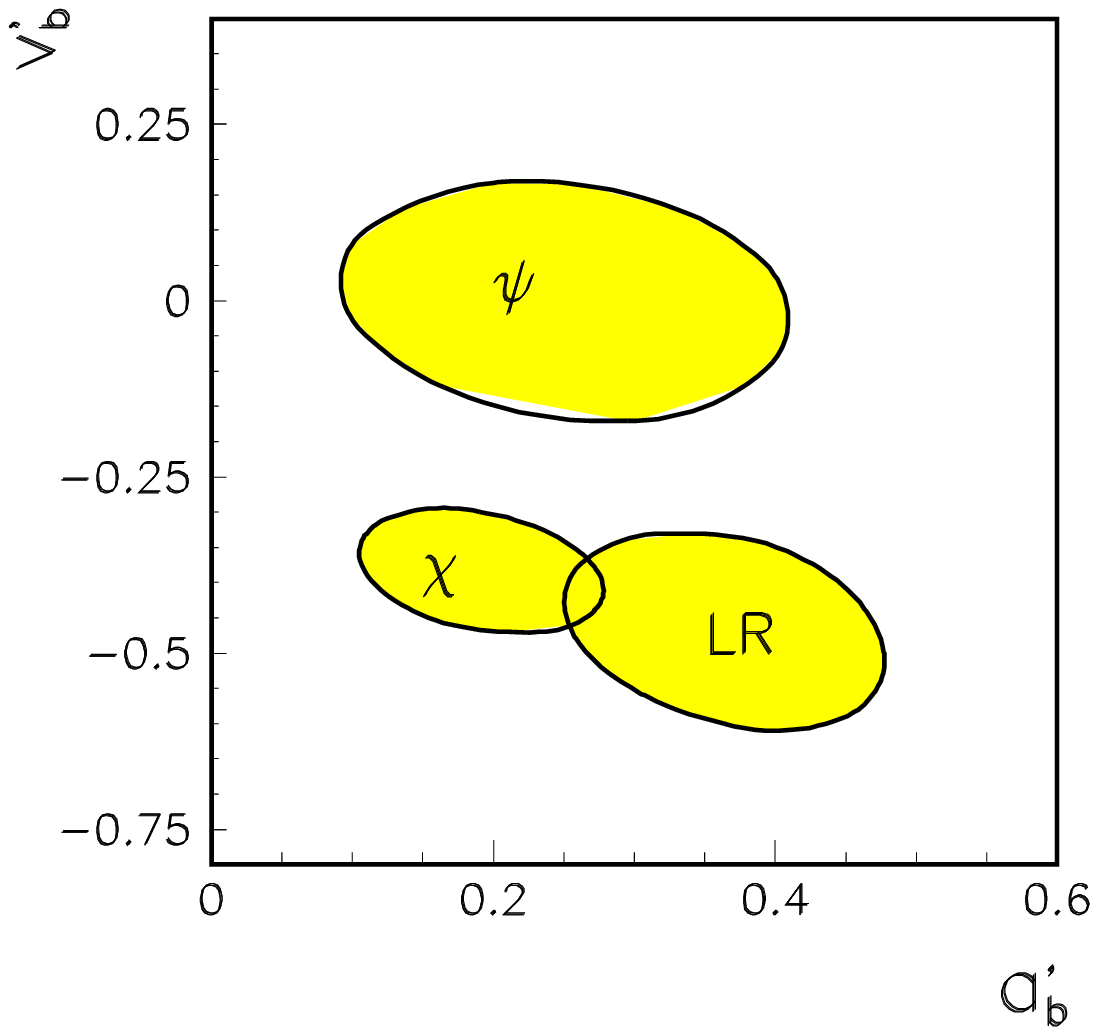}
}
 \end{center}
\caption{Model discrimination for $m_{z'}=1$~TeV studying
$e^+ e^- \rightarrow b \bar b$
at $\sqrt{s}=0.5$~TeV with ${\cal L}= 50$~fb$^{-1}$. 
A systematic error of 1\% for all
b-quark observables is  taken into account.}
\label{fig:6}
\end{figure}

\begin{figure}[htbp]
  \begin{center}
   \mbox{\epsfxsize=8cm\epsffile{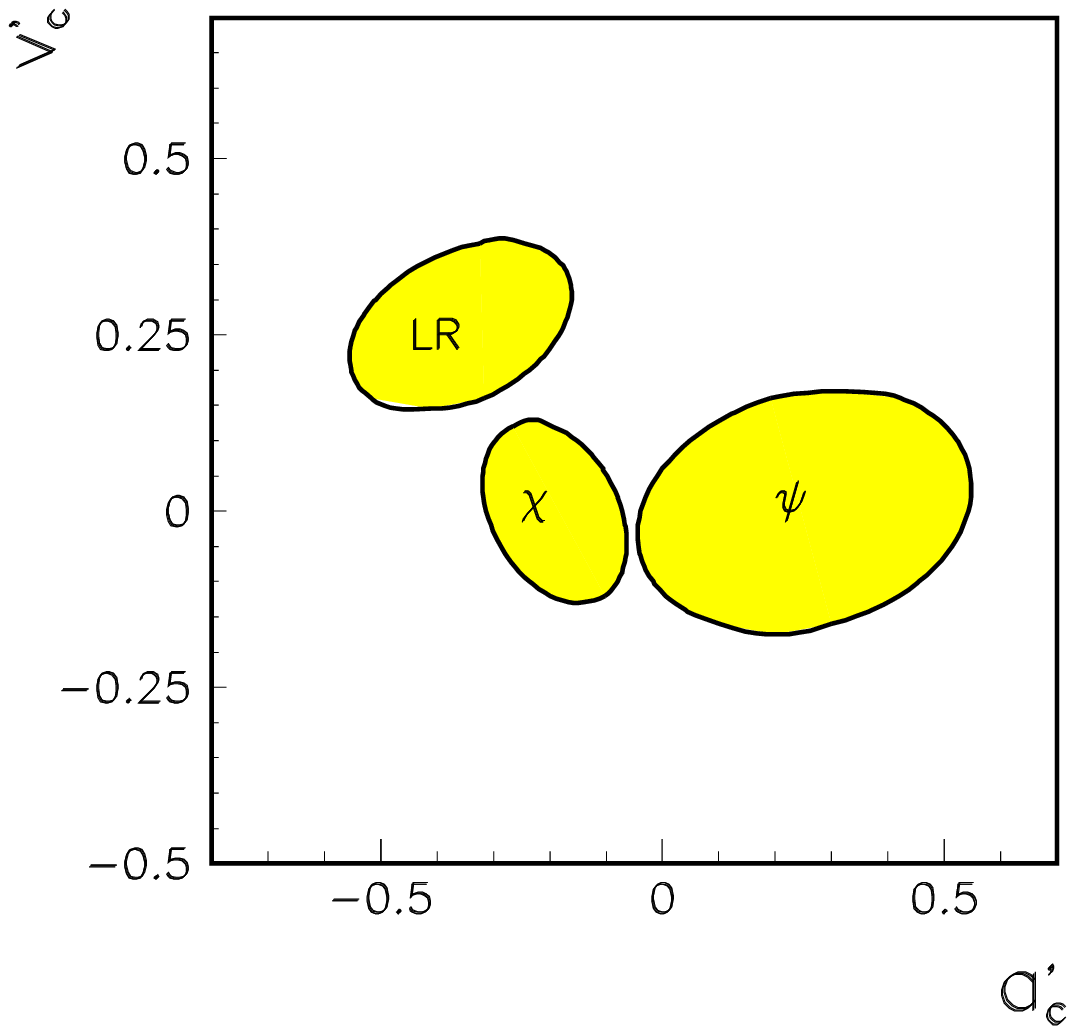}
}
 \end{center}
\caption{Model discrimination for $m_{z'}=1$~TeV studying
$e^+ e^- \rightarrow c \bar c$
at $\sqrt{s}=0.5$~TeV with ${\cal L}= 50$~fb$^{-1}$.
A systematic error of 1.5\% for all
c-quark observables is  taken into account.}
\label{fig:7}
\end{figure}

\begin{figure}[htbp]
  \begin{center}
   \mbox{\epsfxsize=8cm\epsffile{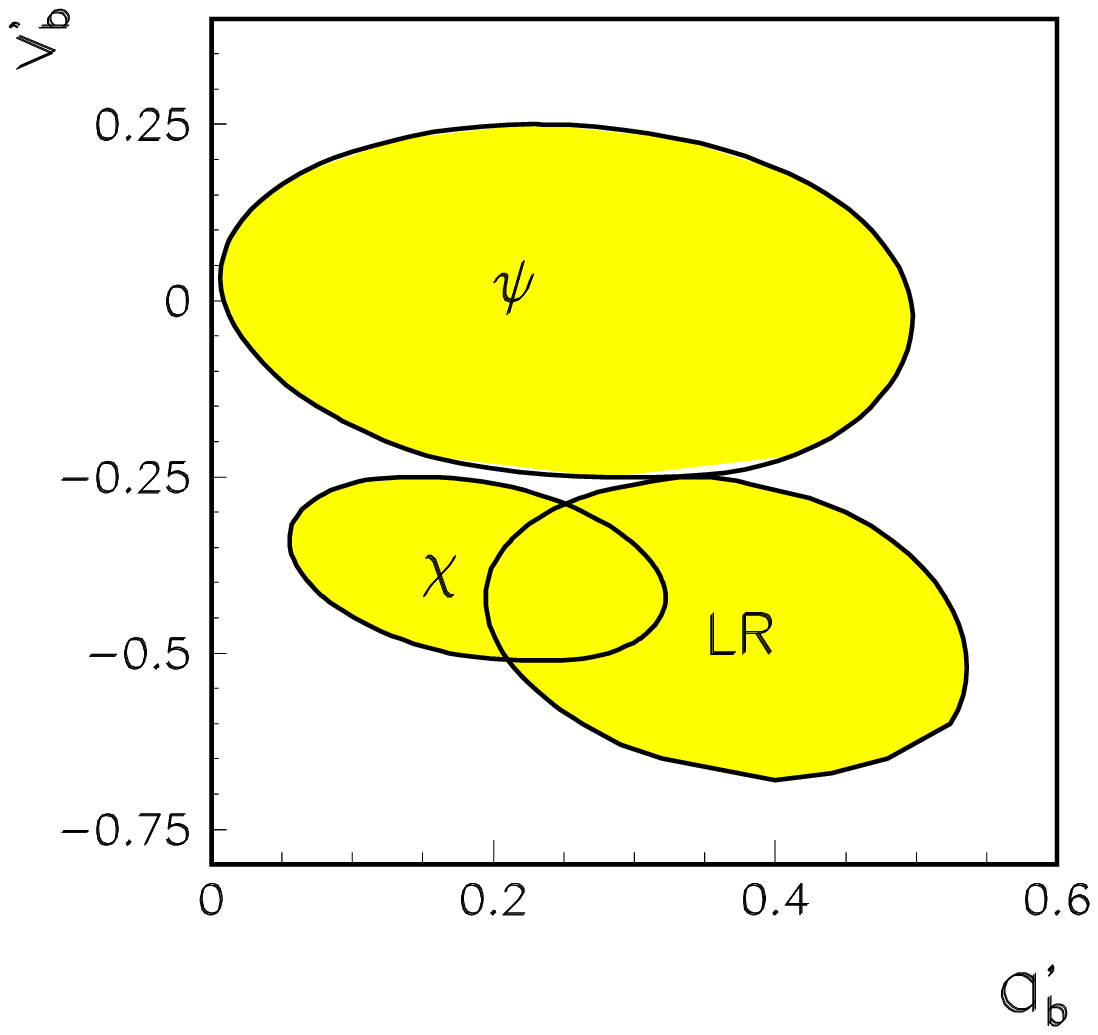}
}
 \end{center}
\caption{Model discrimination for $m_{z'}=3$~TeV studying
$e^+ e^- \rightarrow b \bar b$
at $\sqrt{s}=1.5$~TeV with ${\cal L}=100$~fb$^{-1}$.
A systematic error of 1\% for all
b-quark observables is  taken into account.}
\label{fig:8}
\end{figure}

\begin{figure}[htbp]
  \begin{center}
   \mbox{\epsfxsize=8cm\epsffile{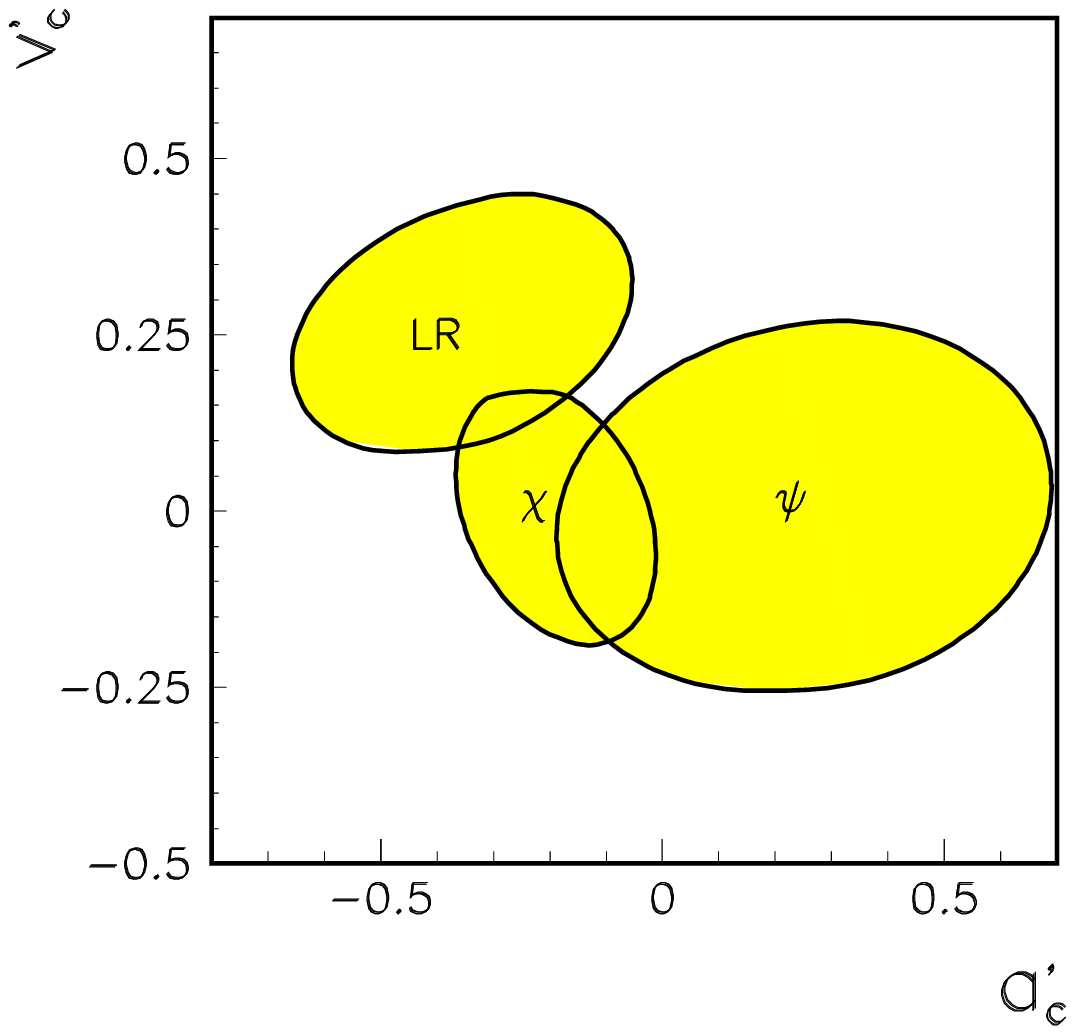}
}
 \end{center}
\caption{Model discrimination for $m_{z'}=3$~TeV studying
$e^+ e^- \rightarrow c \bar c$
at $\sqrt{s}=1.5$~TeV with ${\cal L}=100$~fb$^{-1}$.
A systematic error of 1.5\% for all
c-quark observables is  taken into account.}
\label{fig:9}
\end{figure}

In figure 5  expected results for Z$' b \bar b$ couplings are
shown for different fiducial volumes of the vertex detector.
An ideal  vertex detector leads to only slightly better results.

\section{Model Identification}
Figures 6, 7  and 8, 9 show the expected results for
 Z$' b \bar b$  and  Z$' c \bar c$ couplings for the
two different collider scenarios (\ref{coll} a) and  (\ref{coll} c)
assuming $m_{Z'}= $1~TeV and  $m_{Z'}= $3~TeV, respectively.
It is assumed that a Z$'$ in the $\chi,~\psi$, or LR model exists.
Systematic errors of 1\% for b-quark observables and 1.5\% for c-quark
observables are taken into account.
The figures demonstrate that a Z$'$ model can be separated
although the collider operates
below a Z$'$ resonance -- but only to some extent.
The crucial point is the ratio of Z$'$ mass and centre--of--mass
energy, $R=m_{Z'}/\sqrt{s}$. If $R>2$ the distinction
between most of the models becomes nearly impossible if quark
final states are analyzed.
(see also \cite{al, alsr}).
An increase of luminosity cannot improve this substantially
(see figures 3 and 4)
and the systematic errors must not neglected.
The detection limits are reached.

The situation becomes still worse if the Z$'$ mass is unknown.
In order to perform  a simultaneous fit of  Z$'$ mass and couplings
the available luminosity should be distributed on
different of  centre--of--mass energies since
the contributions
$\sigma_{\gamma Z'}, \sigma_{Z Z'}$ and $\sigma_{Z'Z'}$ to the
observables
vary as a function of $\sqrt{s}$ and $m_{Z'}$.
This couls guarantee
a sufficient number of data points 
and a well-defined $\chi^2$. 
But the distribution of luminosity on several 
centre--of--mass points and
the uncertainty of $m_{z'}$  enlarges the
allowed range for Z$'$ couplings.
Furthermore, if the difference between centre--of--mass energy
and Z$'$ mass is large, the sensitivity for model-independent searches can
be lost.
More details on a simultaneous determination
of $m_{Z'},~a'_f$ and $v'_f$ can be found  in \cite{rizzo_nlc}.

\section{Conclusions}

If a Z$'$ boson with a mass  $m_{Z'} < 6\times \sqrt{s}$
exists 
observables measured at NLC deviate from their
Standard Model expectations.
The interpretation of these deviations within special Z$'$ models
gives the Z$'$ mass.
More interesting is a model-independent analysis.
With the determination of
Z$' f \bar f$ couplings  conclusions on the
Z$'$ model become  possible.
If the Z$'$ mass is known a good separation of Z$'$
models is possible for  $m_{Z'} < 3 \times \sqrt{s}$ from lepton
pair production.
In case of quarkonic final states the accuracy of the Z$' q \bar q$
coupling measurement is diminished by
systematic errors which could reach the
magnitude of the statistical errors. A good model resolution is expected
for  $m_{Z'} < 2 \times \sqrt{s}$ for the considered collider scenarios.

\section*{Acknowledgement}
I would like to thank S. Godfrey, J. Hewett, H. Kagan
and T. Rizzo for discussions stimulating this work.
I am grateful to Arnd Leike for close collaboration.

\end{document}